\makeatletter
\def\thanks#1{\protected@xdef\@thanks{\@thanks
        \protect\footnotetext{#1}}}
\makeatother

\documentclass{article}

\PassOptionsToPackage{square,comma,numbers,sort&compress}{natbib}
\usepackage[preprint]{neurips_2023}

\usepackage{notoccite}

\usepackage[utf8]{inputenc} 
\usepackage[T1]{fontenc}    
\usepackage{hyperref}       
\usepackage{url}            
\usepackage{booktabs}       
\usepackage{amsfonts}       
\usepackage{nicefrac}       
\usepackage{microtype}      
\usepackage{xcolor}         
\usepackage{caption}
\usepackage{subcaption}
\usepackage{graphicx}
\usepackage{amssymb}
\usepackage{import}
\usepackage{algorithm,algorithmicx,algpseudocode}
\usepackage{multirow}
\usepackage{tablefootnote}
\usepackage{pifont}
\usepackage{amsmath}
\usepackage{amsthm} 
\floatstyle{plain}
\restylefloat{figure}

\title{Analyzable Chain-of-Musical-Thought Prompting\\for High-Fidelity Music Generation}

\author{%
  {Max W. Y. Lam}$^{*\dagger}$, Yijin Xing$^{*}$\thanks{$^*$Equal contribution}, Weiya You, Jingcheng Wu, Zongyu Yin,\\
  \textbf{Fuqiang Jiang, Hangyu Liu, Feng Liu, Xingda Li, Wei-Tsung Lu, Hanyu Chen,}\\
  \textbf{Tong Feng, Tianwei Zhao, Chien-Hung Liu, Xuchen Song$^{\dagger}$\thanks{$^{\dagger}$Corresponding author}, Yang Li, Yahui Zhou}\\
  Kunlun Inc.\\
  \texttt{\{maxwy.lam, xuchen.song\}@kunlun-inc.com}\\
}

\begin{document}

\maketitle

\begin{abstract}
Autoregressive (AR) models have demonstrated impressive capabilities in generating high-fidelity music. However, the conventional next-token prediction paradigm in AR models does not align with the human creative process in music composition, potentially compromising the musicality of generated samples.
To overcome this limitation, we introduce MusiCoT, a novel chain-of-thought (CoT) prompting technique tailored for music generation. MusiCoT empowers the AR model to first outline an overall music structure before generating audio tokens, thereby enhancing the coherence and creativity of the resulting compositions. By leveraging the contrastive language-audio pretraining (CLAP) model, we establish a chain of ``musical thoughts'', making MusiCoT scalable and independent of human-labeled data, in contrast to conventional CoT methods.
Moreover, MusiCoT allows for in-depth analysis of music structure, such as instrumental arrangements, and supports music referencing -- accepting variable-length audio inputs as optional style references. This innovative approach effectively addresses copying issues, positioning MusiCoT as a vital practical method for music prompting.
Our experimental results indicate that MusiCoT consistently achieves superior performance across both objective and subjective metrics, producing music quality that rivals state-of-the-art generation models.\\
Our samples are available at \url{https://MusiCoT.github.io/}.
\end{abstract}

\section{Introduction}
In recent years, the field of audio generation has experienced significant advancements \cite{dhariwal2020jukebox,agostinelli2023musiclm,copet2023simple,parker2024stemgen,pasini2022musika,liu2023audioldm, schneider2023mo,huang2023noise2music,chen2024musicldm,li2024jen,evans2024fast,lam2023efficient,bai2024seed,lei2024songcreator} with the emergence of deep generative methods. Despite these developments, the challenge of producing high-fidelity and realistic music remains a formidable task. Music generation requires a delicate balance: integrating vocals with a rich tapestry of instruments while maintaining a coherent melodic and harmonic structure, all while ensuring the accuracy of the linguistic content. Human listeners are particularly attuned to musical dissonance, leaving little room for error in generated compositions. Additionally, creating realistic music demands that models adeptly capture the intricacies of the full frequency spectrum. This challenge is further compounded in long-context music generation, where achieving an optimal balance between generation quality and computational efficiency is crucial.

In the realm of music generation, three primary classes of models have emerged as dominant players: (1) autoregressive (AR) models \cite{agostinelli2023musiclm,copet2023simple,parker2024stemgen,yuan2025yue}, (2) diffusion models \cite{pasini2022musika, schneider2023mo,huang2023noise2music,lam2023efficient,chen2024musicldm,li2024jen,evans2024fast,yuan2025yue}, and (3) a hybrid approach that combines language models (LM) with diffusion models \cite{lam2023efficient, bai2024seed,lei2024songcreator}, referred to as the \textit{MeLoDy} framework as in \cite{lam2023efficient}. Each of these approaches has its own strengths and weaknesses, which are explored in detail in Section \ref{sec:related-work}. This paper focuses on the promising MeLoDy framework, which serves as a backbone for this work. Within this framework, the language model plays a crucial role in aligning compositional inputs -- such as text prompts and lyrics -- with the generated musical content. However, it is important to note that this model operates under the paradigm of next-token prediction, which presents a significant limitation. Unlike the autoregressive nature of the model, the creative process of human music composition is inherently non-autoregressive \cite{burgess2013art}. Music producers typically engage in a thoughtful exploration of various elements -- such as mood, genre, instruments, and melodies -- before finalizing a music piece. Notably, this intermediate reasoning aligns well with the concept of chain-of-thought (CoT) prompting \cite{wei2022chain}, where the LM is trained to generate solutions in a step-by-step manner using natural language. Despite the potential of applying CoT techniques to music AR models, this approach remains largely underexplored in the field of music generation.

\begin{figure}[t]
    \centering
    \includegraphics[width=\linewidth]{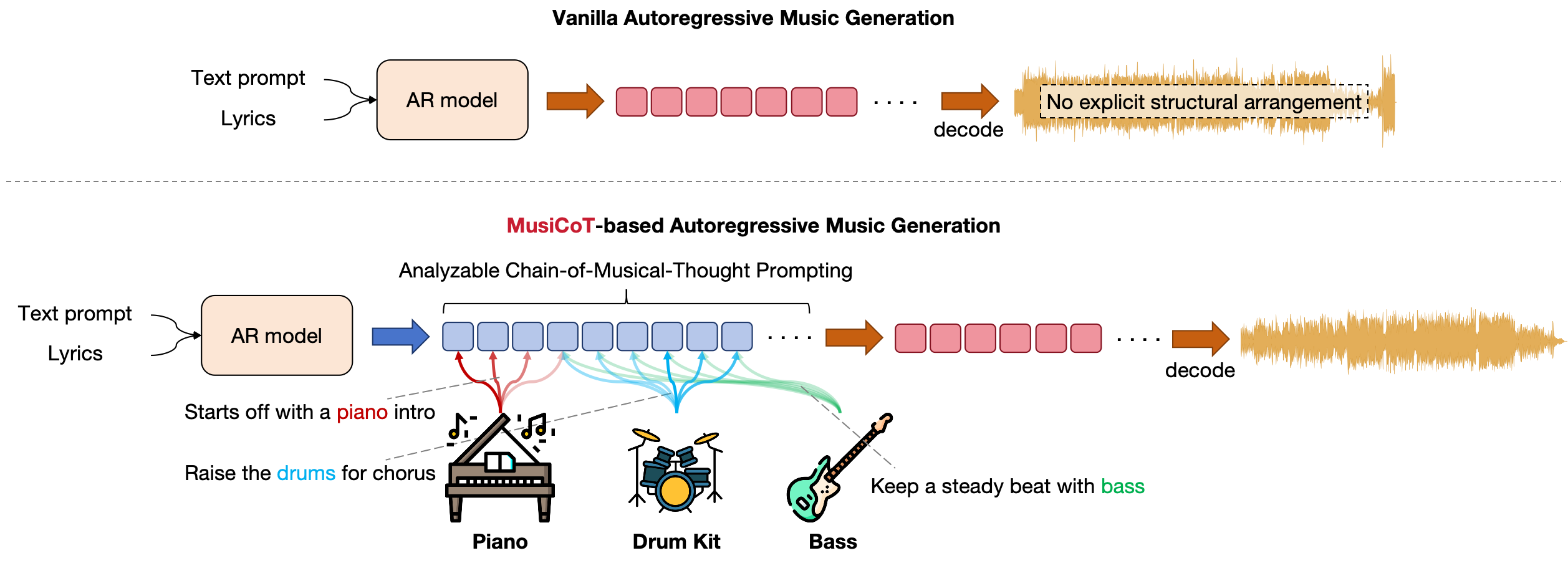}
    \caption{This illustration showcases the MusiCoT reasoning process for music generation, focusing on instrumental arrangement. The arrows are color-coded to indicate the intensity of each instrument: darker colors represent higher intensity, while lighter shades signify lower intensity.}
    \label{fig:1}
\end{figure}

In pursuit of high-fidelity music generation, this paper introduces MusiCoT, an innovative chain-of-thought (CoT) prompting technique specifically designed for music creation. As illustrated in Figure \ref{fig:1}, MusiCoT enables the AR model to first establish a comprehensive and analyzable music structure before generating audio tokens. By leveraging the contrastive language-audio pretraining (CLAP) model \cite{elizalde2023clap}, we define a coherent chain of ``musical thoughts''. The key contributions of MusiCoT are encapsulated in the 4S framework:

\begin{itemize}
    \item \textbf{Scalability}: MusiCoT is built on a separately pretrained CLAP model, allowing for easy scaling with the base AR model without the need for human-labeled data.
    \item \textbf{Structural Analyzability}: In the light of CLAP, MusiCoT offers structural analyzability, facilitating the analysis of elements, e.g., instrumental arrangements, as shown in Figure \ref{fig:1}.
    \item \textbf{Support for Music Reference}: By making slight adjustments to the inference strategy, MusiCoT seamlessly supports music referencing, enabling the input of variable-length audio as an optional style reference. Our experiments demonstrate that MusiCoT effectively mitigates copying issues, making it advantageous for abstractive music referencing.
    \item \textbf{Superior Generation Performance}: Empirical evidence shows that integrating MusiCoT within the MeLoDy framework consistently yields exceptional generation performance, as measured by both objective metrics and subjective evaluations, resulting in music quality that competes with state-of-the-art (SoTA) music generation models.
\end{itemize}

\section{Related Work}
\label{sec:related-work}

\subsection{Music Generation}
In this section, we provide a concise overview of conventional music generation models, highlighting the significant advancements brought about by large language models (LLMs) \cite{achiam2023gpt,grattafiori2024llama,guo2025deepseek} in reasoning capabilities. This has led to the emergence of autoregressive (AR) music generative models \cite{agostinelli2023musiclm,copet2023simple,parker2024stemgen,yuan2025yue}, which have achieved groundbreaking results in the field. However, the performance of these AR-based methods is often limited by the VQ-VAE \cite{dhariwal2020jukebox}, which reconstructs waveforms from a sequence of codes but suffers from low bitrate constraints when utilizing a single codebook \cite{copet2023simple}. To address this limitation, recent works have introduced residual vector quantization (RVQ) techniques \cite{zeghidour2021soundstream, kumar2023high}, enhancing bitrate and promoting the use of multiple codebooks. Nonetheless, the complexity of AR modeling increases significantly when predicting tokens across multiple codebooks \cite{agostinelli2023musiclm, copet2023simple}.

On the other hand, diffusion-based music generative models \cite{pasini2022musika, schneider2023mo,huang2023noise2music,lam2023efficient,chen2024musicldm,li2024jen,evans2024fast,yuan2025yue}, while operating in continuous space \cite{ho2020denoising, rombach2022high} without the drawbacks of quantization loss, demonstrate the potential for superior audio generation quality. However, these models face challenges in supporting long-context windows due to the high computational costs associated with continuous latent vectors \cite{rombach2022high}, which may compromise the musicality and structural coherence of generated samples.

Recognizing the limitations of both approaches, researchers have introduced a cascade model of LMs and diffusion, known as MeLoDy \cite{lam2023efficient}. This framework leverages an LM to predict semantic tokens from a single codebook derived from self-supervised learning (SSL). These tokens then guide a diffusion model for fine-grained waveform generation. As a result, MeLoDy not only produces high-quality music audio comparable to diffusion-based methods but also accommodates long-context windows akin to AR models. This promising cascade approach is rapidly gaining traction in both music generation \cite{bai2024seed,lei2024songcreator} and speech synthesis \cite{anastassiou2024seed,du2024cosyvoice}, showcasing its transformative potential.


\subsection{Chain-of-Thought Prompting}

As a groundbreaking work, \citet{wei2022chain} introduced the concept of chain-of-thought (CoT) prompting, which involves generating a sequence of intermediate reasoning steps -- akin to a chain of thoughts -- within large language models (LLMs). Their findings indicate that CoT significantly enhances the reasoning capabilities of these models, sparking a wave of subsequent research \cite{wang2022self,suzgun2022challenging,lyu2023faithful,zhang2023multimodal,turpin2023language,feng2023towards,hao2024training}. CoT has also proven instrumental in recent industrial advancements, such as OpenAI's O1 \cite{jaech2024openai} and DeepSeek-R1 \cite{guo2025deepseek}, showcasing its practical value in language modeling.

However, training CoT models to articulate intermediate reasoning processes in natural language can be prohibitively expensive, particularly in the context of music generation.\footnote{Manually transcribing the intermediate reasoning involved in the music creative process for CoT prompting can be prohibitively costly, as it necessitates the expertise of qualified music professionals.} Fortunately, \citet{hao2024training} revealed that reasoning within a latent space offers LLMs the flexibility to think without the constraints of language. This notion is supported by neuroimaging research \cite{fedorenko2024language}, which suggests that human language is primarily designed for communication rather than reasoning. This insight aligns with the philosophy of our proposed MusiCoT framework, which similarly emphasizes an abstractive latent space over costly traditional natural-language-based CoT approaches.

Recent studies in the audio domain have also explored the CoT technique \cite{du2024cot,wang2025spark,yuan2025yue}. Notably, YuE \cite{yuan2025yue} presents a CoT method for music generation that focuses on music segments (e.g. verse, chorus, bridge, etc.) \cite{nieto2020audio} extracted by the All-in-one tool \cite{kim2023all}. However, this approach differs fundamentally from our MusiCoT framework, which offers a more nuanced perspective on music generation.

\section{Music Generation Framework}
This section introduces the foundational music generation framework for our model, specifically the \textbf{M}e\textbf{L}o\textbf{D}y (\textbf{M} for music; \textbf{L} for LM; \textbf{D} for diffusion). MeLoDy \cite{lam2023efficient} serves as an efficient alternative to MusicLM \cite{agostinelli2023musiclm}, giving impressive results in music generation. The MeLoDy framework comprises two distinct modeling stages: the semantic stage and the acoustic stage, which are detailed below.

\subsection{Semantic Modeling with a Semantic Language Model}
In this paper, we define the set of conditional inputs for music generation as $\mathbf{C}$. Our focus is on generating music with and without vocals, represented by $\mathbf{C}:=\{\mathbf{c}_\text{lyrics}, \mathbf{c}_\text{text}\}$. Here, $\mathbf{c}_\text{lyrics}$ specifies the lyrics for the vocal parts, while $\mathbf{c}_\text{text}$ serves as a prompt describing the desired characteristics of the music. We extend the original MeLoDy framework, which was limited to non-vocal text-to-music generation, allowing for the optional omission of $\mathbf{c}_\text{lyrics}$ in instrumental music generation.

With this definition of $\mathbf{C}$, a semantic LM, parameterized by $\boldsymbol{\theta}$, addresses the following problem:
\begin{align}
    p_{\boldsymbol{\theta}}(\mathbf{s}_{1:N}|\mathbf{C})=p_{\boldsymbol{\theta}}(\mathbf{s}_{1}|\mathbf{C})\prod_{n=2}^{N}p_{\boldsymbol{\theta}}(\mathbf{s}_{n}|\mathbf{s}_{1:n-1},\mathbf{C}),
\end{align}
where $\mathbf{s}_{1:N}=[\mathbf{s}_1, \ldots, \mathbf{s}_n]$ represents $N$ semantic tokens generated through quantization methods like K-Means \cite{ahmed2020k} or VQ-VAE \cite{van2017neural}, using self-supervised learning (SSL) encoders such as w2v-BERT \cite{chung2021w2v, agostinelli2023musiclm}, Wav2Vec2-Conformer \cite{baevski2020wav2vec, gulati2020conformer, lam2023efficient}, BEST-RQ \cite{chiu2022self, lei2024songcreator}, or fused X-Codec \cite{ye2024codec, yuan2025yue}. The goal of semantic modeling is to capture meaningful representations, including phonemes, melodies, genres, and instruments. Although the MeLoDy framework accommodates various audio tokenizers, the choice of semantic tokenizer is crucial for the performance of the semantic LM and the quality of the final audio. In this study, we adopt the BEST-RQ as our SSL encoder, as detailed in Section \ref{sec:exps}.

\subsection{Acoustic Modeling with a Diffusion Model}
Unlike MusicLM \cite{agostinelli2023musiclm}, which utilizes two distinct LMs for coarse and fine acoustic modeling through multi-codebook prediction,  MeLoDy offers a more streamlined and effective solution by replacing these LMs with a single diffusion model. This diffusion model, parameterized by $\boldsymbol{\phi}$, is designed to recover the true music data distribution  $p_\text{data}(\mathbf{x})$ using a variational distribution $p_{\boldsymbol{\phi}}(\mathbf{x}_0|\mathbf{s}_{1:N})$. It achieves this by executing $T$ steps of reverse process, starting from a prior $p(\mathbf{x}_{T}):=\mathcal{N}(\mathbf{0}, \mathbf{I})$:
\begin{align}
    p_{\boldsymbol{\phi}}(\mathbf{x}_{0}|\mathbf{s}_{1:N})=\mathbb{E}_{\mathbf{x}_{1}, \ldots, \mathbf{x}_{T}}\left[p(\mathbf{x}_{T})\prod_{t=1}^{T}p_{\boldsymbol{\phi}}(\mathbf{x}_{t-1}|\mathbf{x}_{t},\mathbf{s}_{1:N})\right].
\end{align}

Additionally, the authors of MeLoDy \cite{lam2023efficient} introduce a novel dual-path diffusion (DPD) based on an innovative network architecture for the v-diffusion model \cite{salimans2022progressive}. In this paper, we generalize this approach, allowing for flexibility in the diffusion model's architecture, training algorithms, and sampling methods. After reviewing several successful diffusion models in the audio domain, we have chosen the diffusion model presented in Stable Audio \cite{evans2024fast}, which is elaborated upon in Section \ref{sec:exps}.

\subsection{Conditional Music Generation with Lyrics and Text Prompts}

Using text prompts to guide generative models is a common approach in music generation. Previous studies \cite{agostinelli2023musiclm, lam2023efficient, copet2023simple, chen2024musicldm, evans2024fast} typically encode these prompts, denoted as $\mathbf{c}_\text{text}$, into continuous embeddings via a text encoder. However, accurately capturing the intent of these descriptions can be challenging. In our framework, we ensure that the text embedding aligns with the generated semantic tokens by prefixing it to the semantic LM. Notably, MusicLM \cite{agostinelli2023musiclm} and MeLoDy \cite{lam2023efficient} leverage the MuLan model \cite{huang2022mulan}, which jointly embeds music audio and its corresponding text into a joint embedding space, minimizing divergence between them. Similarly, the contrastive language-audio pretraining (CLAP) model \cite{elizalde2023clap} is widely used in text-to-music generation, e.g. in \cite{copet2023simple, chen2024musicldm, evans2024fast}, serving as an alternative to MuLan. During training, we prefix the audio embeddings to the lyrics, and at inference, we can seamlessly switch to using text embeddings for music generation. For optimal performance, we have developed our own CLAP model, which will be detailed in the Appendix.

In addition to cross-domain embedding models, we can utilize outputs from music information retrieval (MIR) models \cite{lerch2022introduction}, which categorize music into tags such as genres, genders, moods, and instruments. This MIR-based prompting method, also employed in YuE \cite{yuan2025yue}, is considered an in-context learning (ICL) approach for music generation. Our findings indicate that combining CLAP-based and MIR-based prompting yields the highest accuracy and quality in generated music samples. Then, the structure of conditions fed into the semantic LM can be formally written as
\begin{align}
    \mathbf{c}_\text{clap} \oplus \text{Emb}_{\boldsymbol\theta}\left(\text{Tokenizer}(\mathbf{c}_\text{tags})\right) \oplus \text{Emb}_{\boldsymbol\theta}\left(\text{Tokenizer}(\mathbf{c}_\text{lyrics})\right),
\end{align}
where $\oplus$ signifies concatenation along the feature dimension. Here, $\text{Emb}_{\boldsymbol\theta}(\cdot)$ refers to the token embedder trained in conjunction with the semantic LM, and $\text{Tokenizer}(\cdot)$ is the method employed for natural language tokenization, such as the subword-based approach using the BERT tokenizer \cite{devlin2019bert}. In addition, we define the components as follows:
\begin{align}
    \mathbf{c}_\text{clap} &= \begin{cases} \text{CLAP}_\text{audio}(\mathbf{x}), & \text{at training time;}\\ \text{CLAP}_\text{text}(\mathbf{c}_\text{text}),& \text{at inference time}, \end{cases}\\
    \mathbf{c}_\text{tags} &= \begin{cases} \text{MIR}(\mathbf{x}), & \text{at training time;}\\
\left\{\forall \mathbf{c}_\text{tag}\in\mathbb{T} \,|\, \frac{\text{CLAP}_\text{text}(\mathbf{c}_\text{text})^\top \text{CLAP}_\text{text}(\mathbf{c}_\text{tag})}{\lVert\text{CLAP}_\text{text}(\mathbf{c}_\text{text})\rVert^2_2 \lVert\text{CLAP}_\text{text}(\mathbf{c}_\text{tag})\rVert^2_2}>\delta\right\},& \text{at inference time}, \end{cases}
\end{align}
where $\mathbf{x}\sim p_\text{data}(\mathbf{x})$ represents the training music audio, $\text{CLAP}_\text{audio}(\cdot)$ and $\text{CLAP}_\text{text}(\cdot)$ denote the audio\footnote{Since the audio encoder convert every 10s of audio into an embedding, we randomly choose one of embeddings at training time to improve robustness.} and text encoders in the CLAP model, respectively. The function $\text{MIR}(\cdot)$ encompasses various music information retrieval classifiers, including those for genre, gender, mood, and instrument. The set $\mathbb{T}$ includes all possible outputs from the $\text{MIR}(\cdot)$ classifiers, and $\delta\in(0, 1)$ serves as a threshold for the cosine similarity between the text prompt and each tag in the set.

\begin{figure}[t]
     \centering
     \includegraphics[width=\textwidth]{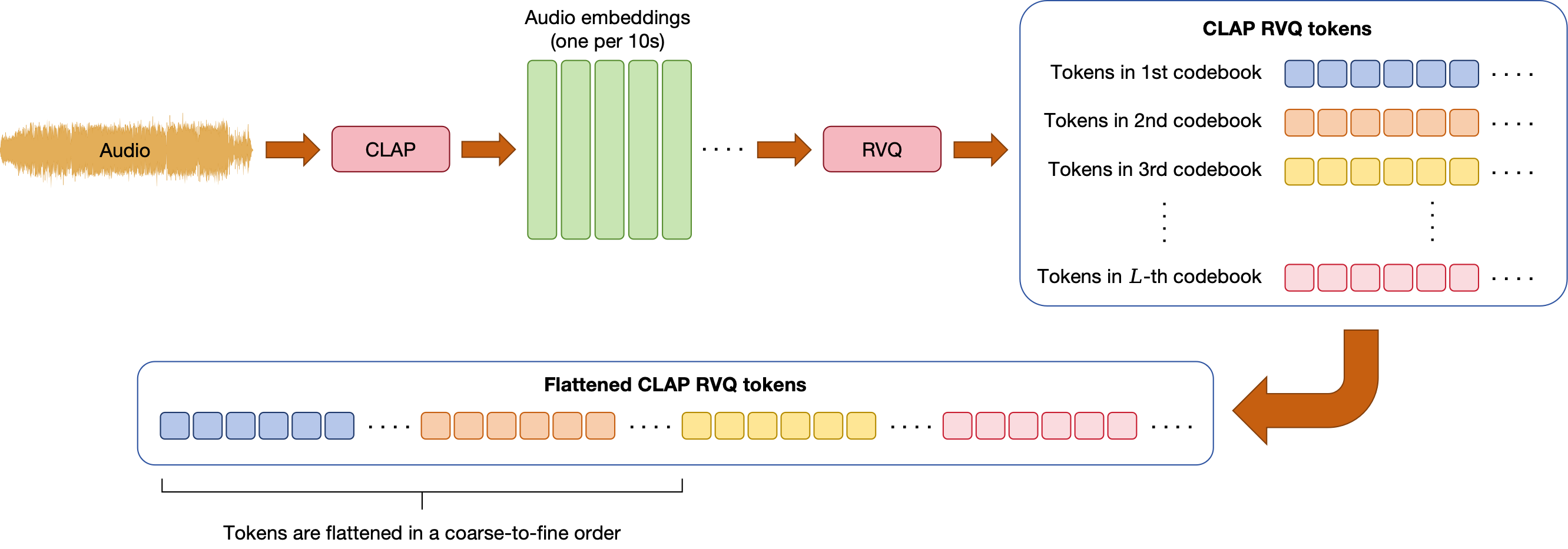}
     \caption{The diagram illustrating the computation of flattened CLAP RVQ tokens given an audio.}
     \label{fig:2}
\end{figure}

\section{MusiCoT: Analyzable Chain-of-Musical-Thought Prompting}
\label{sec:model}

As highlighted in \cite{bartel2006musical}, ``musical thought'' is the foundation of a music producer's creativity. When composing or improvising, producers engage in a distinct decision-making process, effectively ``thinking musically''. This creative journey often involves breaking down the process into intermediate decisions, refining each choice before finalizing the piece. A primary goal of this paper is to equip music generative models with the capability to replicate this chain of musical thought -- creating a coherent series of reasoning and decision-making steps that culminate in a polished music sample.

\subsection{Viewing CLAP Audio Embeddings as Analyzable Musical Thoughts}

This paper proposes a novel approach to representing intermediate musical thoughts using the contrastively trained cross-domain embedding model, known as the CLAP model \cite{elizalde2023clap}, rather than relying on natural language descriptions as seen in \cite{wei2022chain}. The concept of utilizing continuous features is not new; previous research by \citet{hao2024training} indicates that reasoning in a latent space is often more effective than reasoning in natural language. This is further supported by neuro-imaging studies \cite{fedorenko2024language}, which suggest that human language is primarily optimized for communication rather than for reasoning tasks. Specifically, the CLAP model encodes segments of music audio into continuous-valued embeddings every 10 seconds. For a typical 3-minute song, this results in a sequence of audio embeddings, denoted as $\mathbf{C}_\text{clap}:=\left[\mathbf{c}^{(1)}_\text{clap}, \ldots, \mathbf{c}^{(M)}_\text{clap}\right]$. Each embedding, corresponding to a 10-second clip, is analyzable that allows for cosine similarity calculations against any relevant text.

\begin{figure}[t]
     \centering
     \includegraphics[width=\textwidth]{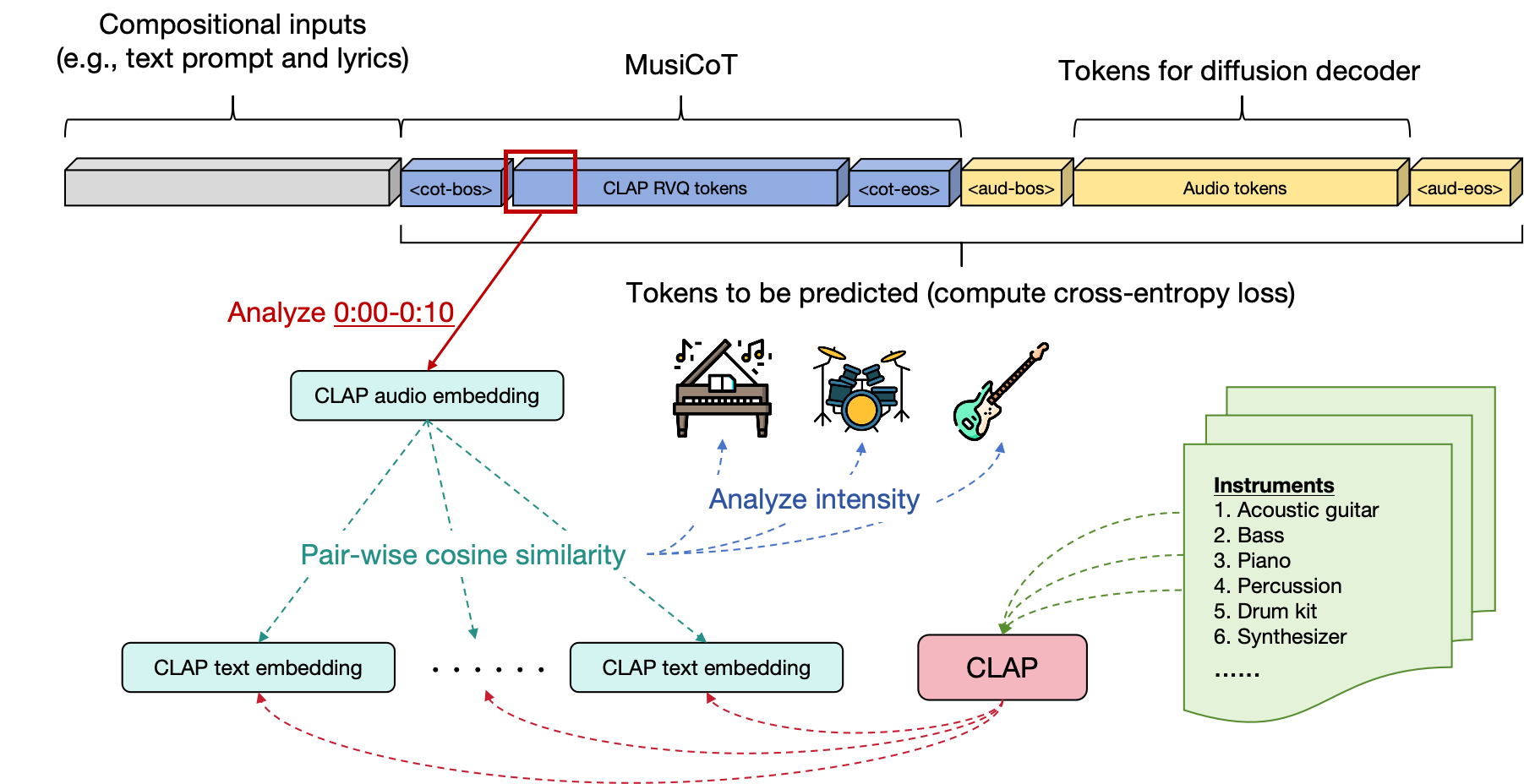}
     \caption{The diagram presenting the token arrangement in MusiCoT-based autoregressive model and the structural analyzability obtained from the CLAP RVQ token prediction.}
     \label{fig:3}
\end{figure}
\subsection{Predicting Coarse-to-Fine Flattened RVQ for a Stabler MusiCoT Training}

With the chain of musical thoughts established, we encounter a significant challenge. The continuous nature of CLAP audio embeddings renders traditional training objectives, such as mean-squared-error (MSE) loss, L1 loss, and contrastive infoNCE loss \cite{oord2018representation}, ineffective for music generation. The aforementioned work \citet{hao2024training} has not explicitly addressed the prediction of continuous thoughts, depends instead on standard CoT training in natural language.

To tackle this issue in MusiCoT, we introduce a residual vector quantization (RVQ) \cite{zeghidour2021soundstream} based coarse-to-fine tokenization method, illustrated in Figure~\ref{fig:2}. This RVQ model consists of $L$ codebooks, parameterized by $\boldsymbol{\zeta}$, and trained using a reconstruction-based quantization loss. Each frame of audio embedding, denoted as $\mathbf{c}^{(m)}_\text{clap}$, is discretized into tokens $c^{(m,1)}_\text{clap}, \ldots, c^{(m,L)}_\text{clap}$, leading to a quantized vector $\tilde{\mathbf{c}}^{(m)}_\text{clap}:=\tilde{\mathbf{c}}^{(m,L)}_\text{clap}$ as defined below: 
\begin{align}
\tilde{\mathbf{c}}^{(m,k)}_\text{clap}=\sum_{k=1}^k\text{Emb}^{(k)}_{\boldsymbol{\zeta}}\left(c^{(m,k)}_\text{clap}\right),\quad c^{(m,k)}_\text{clap} = \underset{q\in\mathbb{Q}^{(k)}}{\arg\min}\,
 \left\lVert\tilde{\mathbf{c}}^{(m, k)}_\text{clap} -\text{Emb}^{(k)}_{\boldsymbol{\zeta}}(q)\right\rVert^2_2
\end{align}
where $\text{Emb}^{(k)}_{\boldsymbol{\zeta}}(\cdot)$ is the $k$-th residual token embedder for the $k$-th codebook, $\tilde{\mathbf{c}}^{(m,k)}_\text{clap}$ denotes the cumulative sum of the first $k$ quantized vectors, and $\mathbb{Q}^{(k)}$ is the index set of the $k$-th codebook.

In MusiCoT, we arrange the RVQ tokens in a flattened coarse-to-fine sequence for LM prediction, ensuring that coarser tokens are predicted before finer ones. Unlike traditional CoT reasoning, which breaks down complex tasks into smaller steps, music generation requires a holistic approach. Our intermediate musical thoughts are designed to maintain this integrity, with each token sequence corresponding directly to the whole generated music with precise time alignment. The $L$ codebooks represent different levels of granularity, making the generation of these intermediate tokens akin to structuring music from a broad to detailed perspective.

During training, the semantic LM utilizes the flattened CLAP RVQ tokens as additional prediction targets, as shown in Figure~\ref{fig:3}. Similar to standard CoT training, these predicted tokens are treated like audio tokens for cross-entropy (CE) loss computation, with the addition of two special tokens -- <cot\_bos> and <cot\_eos> -- to indicate when to transition from generating MusiCoT tokens to audio tokens. The inherent structure of CLAP embeddings allows for the analysis of predicted RVQ tokens within a joint language-audio latent space, enabling us to explore the musical characteristics of each 10-second audio segment. For instance, we can analyze the arrangement of instruments by computing cosine similarities between generated embeddings and the text embeddings of various instruments, providing insights into how different instruments interact over time in the generated music.

\subsection{Dual-Sampling Strategy for MusiCoT}
\label{sec:dual}

In MusiCoT, we integrate tokens from three domains: text tokens, flattened CLAP RVQ tokens, and audio tokens, into a single LM. This raises an important question regarding the sampling strategy: should we use the same sampling approach for both the flattened CLAP RVQ tokens and the audio tokens, or should we adopt different strategies? This issue is relatively unexplored in existing literature. In this section, we present two novel dual-sampling strategies specifically designed for MusiCoT.

\subsubsection{Dual-Temperature Sampling}
A recent study \cite{du2025optimizing} highlights the critical role of temperature selection as a sampling hyperparameter in enhancing language model (LM) performance. Our experimental findings in music generation further support this importance. To leverage this insight, we introduce a dual-temperature sampling method for MusiCoT. This approach involves configuring the semantic LM with two distinct sets of sampling temperatures: one for the flattened CLAP RVQ tokens and another for the audio tokens. The effectiveness of this dual-temperature sampling strategy is demonstrated in Section \ref{sec:exps}.

\subsubsection{Dual-Scale Classifier-Free Guidance}
Classifier-free guidance (CFG) \cite{ho2022classifier} is a versatile technique originally developed for diffusion generative models. Its effectiveness has also been demonstrated in language modeling applications, including AudioGen \cite{kreuk2022audiogen} and MusicGen \cite{copet2023simple}. In our research, we have identified significant benefits of employing CFG within both the semantic language model and the diffusion model, even though the authors of MeLoDy \cite{lam2023efficient} did not explore CFG-based sampling for the semantic LM. For MusiCoT, we introduce a dual-scale CFG sampling strategy that modifies the log probabilities as follows:
\begin{align}
    \log p_{\boldsymbol{\theta}}(\mathbf{c}^{(1:M,1:L)}_\text{clap}|\mathbf{C})&=\lambda_1\log p_{\boldsymbol{\theta}}(\mathbf{c}^{(1:M,1:L)}_\text{clap}|\mathbf{C}) + (1 - \lambda_1)\log p_{\boldsymbol{\theta}}(\mathbf{c}^{(1:M,1:L)}_\text{clap}),\\
    \log p_{\boldsymbol{\theta}}(\mathbf{s}_{1:N}|\mathbf{c}^{(1:M,1:L)}_\text{clap},\mathbf{C})&=\lambda_2\log p_{\boldsymbol{\theta}}(\mathbf{s}_{1:N}|\mathbf{c}^{(1:M,1:L)}_\text{clap},\mathbf{C}) + (1 - \lambda_2)\log p_{\boldsymbol{\theta}}(\mathbf{s}_{1:N}),
\end{align}
where $\mathbf{c}^{(1:M,1:L)}_\text{clap}:=[c^{(1,1)}_\text{clap}, \ldots, c^{(M,1)}_\text{clap}, c^{(1,2)}_\text{clap}, \ldots, c^{(M,2)}_\text{clap}, \ldots, c^{(M,L)}_\text{clap}]$ represents the flattened CLAP RVQ tokens.

\section{Experiments}
\label{sec:exps}

\subsection{Experimental Setup}

\paragraph{Model Setup}

In this work, we employ the LLaMA architecture \cite{touvron2023llama}, as suggested in \cite{lam2023efficient}, for the semantic LM, except opting for a larger variant with approximately 1 billion parameters. For the self-supervised learning (SSL) model that generates audio tokens, we utilize the BEST-RQ model \cite{chiu2022self} with a frame rate of 25Hz. For better pronunciation clarity, the BEST-RQ model is fine-tuned with CTC loss \cite{graves2006connectionist} given paired music audio and lyrics, in a way similar to the approach in \cite{zhang2023google}. Our diffusion model mirrors the architecture, training and sampling pipeline of Stable Audio\footnote{https://github.com/Stability-AI/stable-audio-tools} \cite{kong2021fast}, also featuring a model size of about 1 billion parameters to convert audio tokens into high-quality waveforms. Additionally, we replicate the audio VAE-GAN used in Stable Audio, reconstructing audio at 44.1kHz with 43Hz 64-dim latents. For the CLAP model, we train a music-focused version using the official implementation\footnote{https://github.com/LAION-AI/CLAP}. Following this, we train the RVQ model using a publicly available implementation\footnote{https://github.com/lucidrains/vector-quantize-pytorch}. Detailed configurations can be found in the Appendix.

\paragraph{Data Preparation} 

The models in this paper -- including SSL, CLAP, RVQ, semantic LM, audio VAE-GAN, and diffusion model -- are trained on approximately 10 million English songs sourced from DISCO-10M \cite{lanzendorfer2023disco} and around 200,000 confidential in-house music tracks. Data preprocessing is essential to meet the diverse requirements of each model. Initially, we use the Demucs \cite{defossez2019demucs, defossez2021hybrid, rouard2022hybrid} music source separation model to extract vocals from the songs. Next, an automatic speech recognition (ASR) model transcribes the extracted vocals and provides timestamps for each line of lyrics. We also employ a voice activity detection (VAD) model to identify silent segments. To segment the music into parts like intro, verse, chorus, break, and outro, we utilize the All-in-One model \cite{kim2023all}. For evaluation, we generate 100 sets of English lyrics using ChatGPT \cite{chatgpt}, incorporating various moods and imaginative scenarios. Additionally, we use ChatGPT to create tags that describe genre, sub-genres, gender, mood, instruments, and subjective feelings. To ensure fairness in comparisons, all models are tested with the same pairs of lyrics and prompts.

\begin{table}
  \caption{We compare our base model, MusiCoT, with conventional music generation models, including leading commercial products.\tablefootnote{Notably, previous works in academic research have struggled to match the performance of these commercial offerings, often omitting them due to significant performance gaps.} The asterisk (*) indicates our own implementation of the base model within the MeLoDy framework for music generation.}
  \label{tab:1}
  \centering
  \begin{tabular}{lccccccc}
    \toprule
    \multirow{2}{*}{\textbf{Music Generator}} & \multirow{2}{*}{\textbf{RTF} ($\downarrow$)} & \multirow{2}{*}{\textbf{MOS} ($\uparrow$)} &   \multirow{2}{*}{\textbf{FAD} ($\downarrow$)} &  \multicolumn{4}{c}{\textbf{Content Scores} ($\uparrow$)} \\
    \cmidrule(r){5-8}
    & & & & \textbf{CE} & \textbf{CU} & \textbf{PC} & \textbf{PQ}
    \\
    \midrule
    Suno V4 \cite{sunov4} & 0.84 & \textbf{3.77} & {0.122} & \textbf{7.58} & \textbf{7.87} & 6.28 & \textbf{8.21}\\
    Udio V1.5 \cite{udio1.5} & 1.48 & 3.62 & \textbf{0.110} & 7.14 & 7.29 & \textbf{6.72} & 7.46 \\
    Mureka V5.5 \cite{mureka5.5} & \textbf{0.27} & 3.39 & {0.119} & 7.34 & 7.76 & 6.41 & 8.03 \\
    \midrule
    YuE\tablefootnote{YuE is tested using the model released at {https://github.com/multimodal-art-projection/YuE}} \cite{yuan2025yue} & 12 & 3.00 & 0.287& 7.12 & 7.51 & 5.30 & 7.78 \\
    {*MeLoDy \cite{lam2023efficient}} & \textbf{0.27} & 3.35 & {0.112} & 7.49 & 7.85 & \textbf{6.38} & \textbf{8.11} \\
    *MeLoDy \cite{lam2023efficient} + \textbf{MusiCoT} & \textbf{0.27} & \textbf{3.72} & \textbf{0.102} & \textbf{7.49}&\textbf{7.87}&6.21&\textbf{8.11} \\
    \quad{w/o Dual-Temp.} & \textbf{0.27} & {3.49} & {0.111} & 7.46&7.83&6.18&8.06 \\
    \quad{w/o DS-CFG} & \textbf{0.27}& {3.51} &{0.106} & 7.47&7.84&6.09&8.06\\
    \quad{w/o Dual-Temp. \& DS-CFG} & \textbf{0.27}& {3.48} & {0.113} & 7.41&7.82&6.2&8.03 \\
    \bottomrule
  \end{tabular}
  \vspace{-0.68em}
\end{table}

\paragraph{Evaluation Metrics}
To demonstrate the efficiency of the MeLoDy framework \cite{lam2023efficient}, we evaluate the real-time factor (RTF) -- the time taken to generate one second of audio -- across various commercial products and models running on a NVIDIA RTX 4090 GPU. For subjective assessment, we enlist ten music professionals to rate the overall quality of the generated music on a scale from 1 to 5, focusing on fidelity, musicality, and creativity. The mean opinion score (MOS) is reported for comparative analysis, with detailed evaluation protocols provided in the Appendix. For objective metrics, we utilize the CLAP-based Fréchet audio distance (FAD) \cite{kilgour2019frechet} to measure the fidelity of the generated audio against reference tracks from MUSDB18-HQ \cite{rafii2019musdb18}. It is important to note that while the MusicCaps \cite{agostinelli2023musiclm} test set is commonly used for FAD calculations, the audio quality in MusicCaps, sourced from the Audio Set \cite{gemmeke2017audio}, is significantly inferior to commercially produced music, leading to inaccurate fidelity comparisons -- a concern echoed by the music professionals. Furthermore, previous studies often report VGGish-based FAD, e.g., in \cite{agostinelli2023musiclm, copet2023simple, chen2024musicldm, yuan2025yue}; however, this approach requires downsampling audio to mono 16kHz, which compromises sound quality and renders it less sensitive to quality differences. In contrast, CLAP supports stereo audio at 48kHz, making it more suitable for quality comparisons. Additionally, we employ the Meta Audiobox-Aesthetic \cite{tjandra2025meta}, which leverages advanced neural networks to assess perceived musical aesthetics, including content enjoyment (CE), content usefulness (CU), production complexity (PC), and production quality (PQ).

\subsection{Comparing with the State-of-The-Art Models} 
Table \ref{tab:1} summarizes the results of our comparisons. We first assessed the performance of MusiCoT by contrasting the MeLoDy base model with and without MusiCoT. Notably, incorporating MusiCoT does not increase inference time, as measured by the RTF, yet it significantly boosts the MOS and slightly improves the FAD and content scores. This confirms that MusiCoT effectively enhances music generation performance. We then benchmarked our model against three leading closed-source music generation products: Suno V4 \cite{sunov4}, Udio V1.5 \cite{udio1.5}, and Mureka V5.5 \cite{mureka5.5}. It is important to note that our evaluation, conducted in March 2025, reflects the performance of these products at that specific time due to their black-box nature. As shown in Table \ref{tab:1}, while Mureka V5.5 excels in generation speed, its quality lags behind Suno V4 and Udio V1.5. Suno V4 achieved the highest MOS, indicating superior overall music quality, whereas Udio V1.5 recorded the lowest FAD and higher production complexity (PC), attributed to its exceptional sound quality. Before introducing MusiCoT, our MeLoDy base model lagged behind all commercial products. However, with MusiCoT, we significantly elevate our model's MOS, positioning it as the second-best among all competitors. Furthermore, our MusiCoT-based model achieves the lowest FAD among all competitors, highlighting the high fidelity of the generated music and underscoring the practical value of MusiCoT. We invite you to visit our demo page\footnote{https://MusiCoT.github.io} to experience the enhanced fidelity brought by MusiCoT.

\subsection{Structural Analyzability with MusiCoT}

A key feature of MusiCoT is its capability to analyze intermediate musical content using predicted CLAP RVQ tokens in conjunction with various text anchors. These text anchors are a fixed set of embeddings that represent common music-related terms, such as instrument names. To illustrate this functionality, we utilized the `htdemucs\_6s' model from Demucs \cite{defossez2019demucs, defossez2021hybrid, rouard2022hybrid}, which separates audio into six distinct tracks: vocals, bass, drums, piano, guitar, and other accompaniment. We selected five text anchors --`vocals', `bass', `drums', `guitar', and `piano' -- each defined by unique musical characteristics. By calculating the cosine similarities between the quantized CLAP embeddings derived from the generated RVQ tokens and these text anchors, we assessed the correlation with the corresponding track volumes using the Pearson correlation coefficient ($r$). Over 30-second segments, we found the average correlation coefficients for each text anchor: $r_\text{vocals}=0.689$, $r_\text{bass}=0.584$, $r_\text{drums}=0.639$, $r_\text{guitar}=0.628$, and $r_\text{piano}=0.531$. To neglect noise from the music separator, we excluded any track with an average volume below $10^{-2}$ from our calculations. Notably, these results reveal a positive correlation between the cosine similarities and track volumes, reinforcing our hypothesis regarding the analyzability of MusiCoT.

\subsection{Music Referencing with MusiCoT}
One of the standout features of MusiCoT is its seamless support for music referencing. By extracting CLAP RVQ tokens from a reference track, MusiCoT can efficiently transition to predicting audio tokens. This capability enhances music generation by allowing variable-length audio inputs, unlike traditional CLAP-based methods that are limited to fixed lengths, as noted in \cite{lam2023efficient}. Additionally, while continuation-based strategies (such as ICL \cite{yuan2025yue}) offer another approach to audio referencing, they carry a significant risk of the model replicating token sequences from training data. MusiCoT addresses this concern effectively, as CLAP RVQ tokens are more abstract and less prone to copying than standard audio tokens. For practical examples of music referencing, please visit our demo page.

\subsection{Ablation study on Dual-Sampling Strategies}
In this section, we explore the impact of two dual-sampling strategies: dual-temperature sampling (Dual-Temp.) and dual-scale classifier-free guidance (DS-CFG), as introduced in Section \ref{sec:dual}. For Dual-Temp., we set the temperature for MusiCoT tokens at $0.65$ and for audio tokens at $0.75$. In the case of DS-CFG, we use $\lambda_1=2.3,\lambda_2=1.3$, determined through a grid search algorithm. Our findings, presented in lower part of Table \ref{tab:1}, indicate that omitting either or both strategies results in a decline across all performance metrics. Notably, the combined application of Dual-Temp. and DS-CFG yields a more substantial enhancement in performance than utilizing either strategy alone.


\section{Conclusion}
In conclusion, this paper presents MusiCoT, a novel chain-of-thought prompting technique that enhances high-fidelity music generation by aligning the creative processes of AR models with musical thought. By leveraging the CLAP model, MusiCoT not only ensures structural analyzability but also supports music referencing, thereby elevating the quality of generated music. Our experimental results demonstrate that MusiCoT consistently leads to superior generation performances, establishing itself as a valuable advancement in music generation. Ultimately, MusiCoT paves the way for future explorations of generative AI, merging artificial intelligence with the artistry of human creativity.

\medskip

\bibliographystyle{unsrtnat}
\bibliography{neurips_2023}

\begin{thebibliography}{72}
\providecommand{\natexlab}[1]{#1}
\providecommand{\url}[1]{\texttt{#1}}
\expandafter\ifx\csname urlstyle\endcsname\relax
  \providecommand{\doi}[1]{doi: #1}\else
  \providecommand{\doi}{doi: \begingroup \urlstyle{rm}\Url}\fi

\bibitem[Dhariwal et~al.(2020)Dhariwal, Jun, Payne, Kim, Radford, and Sutskever]{dhariwal2020jukebox}
Prafulla Dhariwal, Heewoo Jun, Christine Payne, Jong~Wook Kim, Alec Radford, and Ilya Sutskever.
\newblock Jukebox: A generative model for music.
\newblock \emph{arXiv preprint arXiv:2005.00341}, 2020.

\bibitem[Agostinelli et~al.(2023)Agostinelli, Denk, Borsos, Engel, Verzetti, Caillon, Huang, Jansen, Roberts, Tagliasacchi, et~al.]{agostinelli2023musiclm}
Andrea Agostinelli, Timo~I Denk, Zal{\'a}n Borsos, Jesse Engel, Mauro Verzetti, Antoine Caillon, Qingqing Huang, Aren Jansen, Adam Roberts, Marco Tagliasacchi, et~al.
\newblock Musiclm: Generating music from text.
\newblock \emph{arXiv preprint arXiv:2301.11325}, 2023.

\bibitem[Copet et~al.(2023)Copet, Kreuk, Gat, Remez, Kant, Synnaeve, Adi, and D{\'e}fossez]{copet2023simple}
Jade Copet, Felix Kreuk, Itai Gat, Tal Remez, David Kant, Gabriel Synnaeve, Yossi Adi, and Alexandre D{\'e}fossez.
\newblock Simple and controllable music generation.
\newblock \emph{Advances in Neural Information Processing Systems}, 36:\penalty0 47704--47720, 2023.

\bibitem[Parker et~al.(2024)Parker, Spijkervet, Kosta, Yesiler, Kuznetsov, Wang, Avent, Chen, and Le]{parker2024stemgen}
Julian~D Parker, Janne Spijkervet, Katerina Kosta, Furkan Yesiler, Boris Kuznetsov, Ju-Chiang Wang, Matt Avent, Jitong Chen, and Duc Le.
\newblock Stemgen: A music generation model that listens.
\newblock In \emph{ICASSP 2024-2024 IEEE International Conference on Acoustics, Speech and Signal Processing (ICASSP)}, pages 1116--1120. IEEE, 2024.

\bibitem[Pasini and Schl{\"u}ter(2022)]{pasini2022musika}
Marco Pasini and Jan Schl{\"u}ter.
\newblock Musika! fast infinite waveform music generation.
\newblock \emph{arXiv preprint arXiv:2208.08706}, 2022.

\bibitem[Liu et~al.(2023)Liu, Chen, Yuan, Mei, Liu, Mandic, Wang, and Plumbley]{liu2023audioldm}
Haohe Liu, Zehua Chen, Yi~Yuan, Xinhao Mei, Xubo Liu, Danilo Mandic, Wenwu Wang, and Mark~D Plumbley.
\newblock Audioldm: Text-to-audio generation with latent diffusion models.
\newblock \emph{arXiv preprint arXiv:2301.12503}, 2023.

\bibitem[Schneider et~al.(2023)Schneider, Jin, and Sch{\"o}lkopf]{schneider2023mo}
Flavio Schneider, Zhijing Jin, and Bernhard Sch{\"o}lkopf.
\newblock Mo\^{u}sai: Text-to-music generation with long-context latent diffusion.
\newblock \emph{arXiv preprint arXiv:2301.11757}, 2023.

\bibitem[Huang et~al.(2023)Huang, Park, Wang, Denk, Ly, Chen, Zhang, Zhang, Yu, Frank, et~al.]{huang2023noise2music}
Qingqing Huang, Daniel~S Park, Tao Wang, Timo~I Denk, Andy Ly, Nanxin Chen, Zhengdong Zhang, Zhishuai Zhang, Jiahui Yu, Christian Frank, et~al.
\newblock Noise2music: Text-conditioned music generation with diffusion models.
\newblock \emph{arXiv preprint arXiv:2302.03917}, 2023.

\bibitem[Chen et~al.(2024)Chen, Wu, Liu, Nezhurina, Berg-Kirkpatrick, and Dubnov]{chen2024musicldm}
Ke~Chen, Yusong Wu, Haohe Liu, Marianna Nezhurina, Taylor Berg-Kirkpatrick, and Shlomo Dubnov.
\newblock Musicldm: Enhancing novelty in text-to-music generation using beat-synchronous mixup strategies.
\newblock In \emph{ICASSP 2024-2024 IEEE International Conference on Acoustics, Speech and Signal Processing (ICASSP)}, pages 1206--1210. IEEE, 2024.

\bibitem[Li et~al.(2024)Li, Chen, Yao, Wang, Wang, and Wang]{li2024jen}
Peike~Patrick Li, Boyu Chen, Yao Yao, Yikai Wang, Allen Wang, and Alex Wang.
\newblock Jen-1: Text-guided universal music generation with omnidirectional diffusion models.
\newblock In \emph{2024 IEEE Conference on Artificial Intelligence (CAI)}, pages 762--769. IEEE, 2024.

\bibitem[Evans et~al.(2024)Evans, Carr, Taylor, Hawley, and Pons]{evans2024fast}
Zach Evans, CJ~Carr, Josiah Taylor, Scott~H Hawley, and Jordi Pons.
\newblock Fast timing-conditioned latent audio diffusion.
\newblock In \emph{Forty-first International Conference on Machine Learning}, 2024.

\bibitem[Lam et~al.(2023)Lam, Tian, Li, Yin, Feng, Tu, Ji, Xia, Ma, Song, et~al.]{lam2023efficient}
Max~WY Lam, Qiao Tian, Tang Li, Zongyu Yin, Siyuan Feng, Ming Tu, Yuliang Ji, Rui Xia, Mingbo Ma, Xuchen Song, et~al.
\newblock Efficient neural music generation.
\newblock \emph{Advances in Neural Information Processing Systems}, 36:\penalty0 17450--17463, 2023.

\bibitem[Bai et~al.(2024)Bai, Chen, Chen, Chen, Deng, Dong, Hantrakul, Hao, Huang, Huang, et~al.]{bai2024seed}
Ye~Bai, Haonan Chen, Jitong Chen, Zhuo Chen, Yi~Deng, Xiaohong Dong, Lamtharn Hantrakul, Weituo Hao, Qingqing Huang, Zhongyi Huang, et~al.
\newblock Seed-music: A unified framework for high quality and controlled music generation.
\newblock \emph{arXiv preprint arXiv:2409.09214}, 2024.

\bibitem[Lei et~al.(2024)Lei, Zhou, Tang, Lam, Liu, Wu, Kang, Wu, Meng, et~al.]{lei2024songcreator}
Shun Lei, Yixuan Zhou, Boshi Tang, Max~WY Lam, Hangyu Liu, Jingcheng Wu, Shiyin Kang, Zhiyong Wu, Helen Meng, et~al.
\newblock Songcreator: Lyrics-based universal song generation.
\newblock \emph{Advances in Neural Information Processing Systems}, 37:\penalty0 80107--80140, 2024.

\bibitem[Yuan et~al.(2025)Yuan, Lin, Guo, Zhang, Pan, Zang, Liu, Liang, Ma, Du, et~al.]{yuan2025yue}
Ruibin Yuan, Hanfeng Lin, Shuyue Guo, Ge~Zhang, Jiahao Pan, Yongyi Zang, Haohe Liu, Yiming Liang, Wenye Ma, Xingjian Du, et~al.
\newblock Yue: Scaling open foundation models for long-form music generation.
\newblock \emph{arXiv preprint arXiv:2503.08638}, 2025.

\bibitem[Burgess(2013)]{burgess2013art}
Richard~James Burgess.
\newblock \emph{The art of music production: The theory and practice}.
\newblock Oxford University Press, 2013.

\bibitem[Wei et~al.(2022)Wei, Wang, Schuurmans, Bosma, Xia, Chi, Le, Zhou, et~al.]{wei2022chain}
Jason Wei, Xuezhi Wang, Dale Schuurmans, Maarten Bosma, Fei Xia, Ed~Chi, Quoc~V Le, Denny Zhou, et~al.
\newblock Chain-of-thought prompting elicits reasoning in large language models.
\newblock \emph{Advances in neural information processing systems}, 35:\penalty0 24824--24837, 2022.

\bibitem[Elizalde et~al.(2023)Elizalde, Deshmukh, Al~Ismail, and Wang]{elizalde2023clap}
Benjamin Elizalde, Soham Deshmukh, Mahmoud Al~Ismail, and Huaming Wang.
\newblock Clap learning audio concepts from natural language supervision.
\newblock In \emph{ICASSP 2023-2023 IEEE International Conference on Acoustics, Speech and Signal Processing (ICASSP)}, pages 1--5. IEEE, 2023.

\bibitem[Achiam et~al.(2023)Achiam, Adler, Agarwal, Ahmad, Akkaya, Aleman, Almeida, Altenschmidt, Altman, Anadkat, et~al.]{achiam2023gpt}
Josh Achiam, Steven Adler, Sandhini Agarwal, Lama Ahmad, Ilge Akkaya, Florencia~Leoni Aleman, Diogo Almeida, Janko Altenschmidt, Sam Altman, Shyamal Anadkat, et~al.
\newblock Gpt-4 technical report.
\newblock \emph{arXiv preprint arXiv:2303.08774}, 2023.

\bibitem[Grattafiori et~al.(2024)Grattafiori, Dubey, Jauhri, Pandey, Kadian, Al-Dahle, Letman, Mathur, Schelten, Vaughan, et~al.]{grattafiori2024llama}
Aaron Grattafiori, Abhimanyu Dubey, Abhinav Jauhri, Abhinav Pandey, Abhishek Kadian, Ahmad Al-Dahle, Aiesha Letman, Akhil Mathur, Alan Schelten, Alex Vaughan, et~al.
\newblock The llama 3 herd of models.
\newblock \emph{arXiv preprint arXiv:2407.21783}, 2024.

\bibitem[Guo et~al.(2025)Guo, Yang, Zhang, Song, Zhang, Xu, Zhu, Ma, Wang, Bi, et~al.]{guo2025deepseek}
Daya Guo, Dejian Yang, Haowei Zhang, Junxiao Song, Ruoyu Zhang, Runxin Xu, Qihao Zhu, Shirong Ma, Peiyi Wang, Xiao Bi, et~al.
\newblock Deepseek-r1: Incentivizing reasoning capability in llms via reinforcement learning.
\newblock \emph{arXiv preprint arXiv:2501.12948}, 2025.

\bibitem[Zeghidour et~al.(2021)Zeghidour, Luebs, Omran, Skoglund, and Tagliasacchi]{zeghidour2021soundstream}
Neil Zeghidour, Alejandro Luebs, Ahmed Omran, Jan Skoglund, and Marco Tagliasacchi.
\newblock Soundstream: An end-to-end neural audio codec.
\newblock \emph{IEEE/ACM Transactions on Audio, Speech, and Language Processing}, 30:\penalty0 495--507, 2021.

\bibitem[Kumar et~al.(2023)Kumar, Seetharaman, Luebs, Kumar, and Kumar]{kumar2023high}
Rithesh Kumar, Prem Seetharaman, Alejandro Luebs, Ishaan Kumar, and Kundan Kumar.
\newblock High-fidelity audio compression with improved rvqgan.
\newblock \emph{Advances in Neural Information Processing Systems}, 36:\penalty0 27980--27993, 2023.

\bibitem[Ho et~al.(2020)Ho, Jain, and Abbeel]{ho2020denoising}
Jonathan Ho, Ajay Jain, and Pieter Abbeel.
\newblock Denoising diffusion probabilistic models.
\newblock \emph{Advances in Neural Information Processing Systems}, 33:\penalty0 6840--6851, 2020.

\bibitem[Rombach et~al.(2022)Rombach, Blattmann, Lorenz, Esser, and Ommer]{rombach2022high}
Robin Rombach, Andreas Blattmann, Dominik Lorenz, Patrick Esser, and Bj{\"o}rn Ommer.
\newblock High-resolution image synthesis with latent diffusion models.
\newblock In \emph{Proceedings of the IEEE/CVF Conference on Computer Vision and Pattern Recognition}, pages 10684--10695, 2022.

\bibitem[Anastassiou et~al.(2024)Anastassiou, Chen, Chen, Chen, Chen, Chen, Cong, Deng, Ding, Gao, et~al.]{anastassiou2024seed}
Philip Anastassiou, Jiawei Chen, Jitong Chen, Yuanzhe Chen, Zhuo Chen, Ziyi Chen, Jian Cong, Lelai Deng, Chuang Ding, Lu~Gao, et~al.
\newblock Seed-tts: A family of high-quality versatile speech generation models.
\newblock \emph{arXiv preprint arXiv:2406.02430}, 2024.

\bibitem[Du et~al.(2024{\natexlab{a}})Du, Chen, Zhang, Hu, Lu, Yang, Hu, Zheng, Gu, Ma, et~al.]{du2024cosyvoice}
Zhihao Du, Qian Chen, Shiliang Zhang, Kai Hu, Heng Lu, Yexin Yang, Hangrui Hu, Siqi Zheng, Yue Gu, Ziyang Ma, et~al.
\newblock Cosyvoice: A scalable multilingual zero-shot text-to-speech synthesizer based on supervised semantic tokens.
\newblock \emph{arXiv preprint arXiv:2407.05407}, 2024{\natexlab{a}}.

\bibitem[Wang et~al.(2022)Wang, Wei, Schuurmans, Le, Chi, Narang, Chowdhery, and Zhou]{wang2022self}
Xuezhi Wang, Jason Wei, Dale Schuurmans, Quoc Le, Ed~Chi, Sharan Narang, Aakanksha Chowdhery, and Denny Zhou.
\newblock Self-consistency improves chain of thought reasoning in language models.
\newblock \emph{arXiv preprint arXiv:2203.11171}, 2022.

\bibitem[Suzgun et~al.(2022)Suzgun, Scales, Sch{\"a}rli, Gehrmann, Tay, Chung, Chowdhery, Le, Chi, Zhou, et~al.]{suzgun2022challenging}
Mirac Suzgun, Nathan Scales, Nathanael Sch{\"a}rli, Sebastian Gehrmann, Yi~Tay, Hyung~Won Chung, Aakanksha Chowdhery, Quoc~V Le, Ed~H Chi, Denny Zhou, et~al.
\newblock Challenging big-bench tasks and whether chain-of-thought can solve them.
\newblock \emph{arXiv preprint arXiv:2210.09261}, 2022.

\bibitem[Lyu et~al.(2023)Lyu, Havaldar, Stein, Zhang, Rao, Wong, Apidianaki, and Callison-Burch]{lyu2023faithful}
Qing Lyu, Shreya Havaldar, Adam Stein, Li~Zhang, Delip Rao, Eric Wong, Marianna Apidianaki, and Chris Callison-Burch.
\newblock Faithful chain-of-thought reasoning.
\newblock In \emph{The 13th International Joint Conference on Natural Language Processing and the 3rd Conference of the Asia-Pacific Chapter of the Association for Computational Linguistics (IJCNLP-AACL 2023)}, 2023.

\bibitem[Zhang et~al.(2023{\natexlab{a}})Zhang, Zhang, Li, Zhao, Karypis, and Smola]{zhang2023multimodal}
Zhuosheng Zhang, Aston Zhang, Mu~Li, Hai Zhao, George Karypis, and Alex Smola.
\newblock Multimodal chain-of-thought reasoning in language models.
\newblock \emph{arXiv preprint arXiv:2302.00923}, 2023{\natexlab{a}}.

\bibitem[Turpin et~al.(2023)Turpin, Michael, Perez, and Bowman]{turpin2023language}
Miles Turpin, Julian Michael, Ethan Perez, and Samuel Bowman.
\newblock Language models don't always say what they think: Unfaithful explanations in chain-of-thought prompting.
\newblock \emph{Advances in Neural Information Processing Systems}, 36:\penalty0 74952--74965, 2023.

\bibitem[Feng et~al.(2023)Feng, Zhang, Gu, Ye, He, and Wang]{feng2023towards}
Guhao Feng, Bohang Zhang, Yuntian Gu, Haotian Ye, Di~He, and Liwei Wang.
\newblock Towards revealing the mystery behind chain of thought: a theoretical perspective.
\newblock \emph{Advances in Neural Information Processing Systems}, 36:\penalty0 70757--70798, 2023.

\bibitem[Hao et~al.(2024)Hao, Sukhbaatar, Su, Li, Hu, Weston, and Tian]{hao2024training}
Shibo Hao, Sainbayar Sukhbaatar, DiJia Su, Xian Li, Zhiting Hu, Jason Weston, and Yuandong Tian.
\newblock Training large language models to reason in a continuous latent space.
\newblock \emph{arXiv preprint arXiv:2412.06769}, 2024.

\bibitem[Jaech et~al.(2024)Jaech, Kalai, Lerer, Richardson, El-Kishky, Low, Helyar, Madry, Beutel, Carney, et~al.]{jaech2024openai}
Aaron Jaech, Adam Kalai, Adam Lerer, Adam Richardson, Ahmed El-Kishky, Aiden Low, Alec Helyar, Aleksander Madry, Alex Beutel, Alex Carney, et~al.
\newblock Openai o1 system card.
\newblock \emph{arXiv preprint arXiv:2412.16720}, 2024.

\bibitem[Fedorenko et~al.(2024)Fedorenko, Piantadosi, and Gibson]{fedorenko2024language}
Evelina Fedorenko, Steven~T Piantadosi, and Edward~AF Gibson.
\newblock Language is primarily a tool for communication rather than thought.
\newblock \emph{Nature}, 630\penalty0 (8017):\penalty0 575--586, 2024.

\bibitem[Du et~al.(2024{\natexlab{b}})Du, Ma, Yang, Deng, Chen, Yang, Xiang, Liu, and Qin]{du2024cot}
Yexing Du, Ziyang Ma, Yifan Yang, Keqi Deng, Xie Chen, Bo~Yang, Yang Xiang, Ming Liu, and Bing Qin.
\newblock Cot-st: Enhancing llm-based speech translation with multimodal chain-of-thought.
\newblock \emph{arXiv preprint arXiv:2409.19510}, 2024{\natexlab{b}}.

\bibitem[Wang et~al.(2025)Wang, Jiang, Ma, Zhang, Liu, Li, Liang, Zheng, Wang, Feng, et~al.]{wang2025spark}
Xinsheng Wang, Mingqi Jiang, Ziyang Ma, Ziyu Zhang, Songxiang Liu, Linqin Li, Zheng Liang, Qixi Zheng, Rui Wang, Xiaoqin Feng, et~al.
\newblock Spark-tts: An efficient llm-based text-to-speech model with single-stream decoupled speech tokens.
\newblock \emph{arXiv preprint arXiv:2503.01710}, 2025.

\bibitem[Nieto et~al.(2020)Nieto, Mysore, Wang, Smith, Schl{\"u}ter, Grill, and McFee]{nieto2020audio}
Oriol Nieto, Gautham~J Mysore, Cheng-i Wang, Jordan~BL Smith, Jan Schl{\"u}ter, Thomas Grill, and Brian McFee.
\newblock Audio-based music structure analysis: Current trends, open challenges, and applications.
\newblock \emph{Transactions of the International Society for Music Information Retrieval}, 3\penalty0 (1), 2020.

\bibitem[Kim and Nam(2023)]{kim2023all}
Taejun Kim and Juhan Nam.
\newblock All-in-one metrical and functional structure analysis with neighborhood attentions on demixed audio.
\newblock In \emph{2023 IEEE Workshop on Applications of Signal Processing to Audio and Acoustics (WASPAA)}, pages 1--5. IEEE, 2023.

\bibitem[Ahmed et~al.(2020)Ahmed, Seraj, and Islam]{ahmed2020k}
Mohiuddin Ahmed, Raihan Seraj, and Syed Mohammed~Shamsul Islam.
\newblock The k-means algorithm: A comprehensive survey and performance evaluation.
\newblock \emph{Electronics}, 9\penalty0 (8):\penalty0 1295, 2020.

\bibitem[Van Den~Oord et~al.(2017)Van Den~Oord, Vinyals, et~al.]{van2017neural}
Aaron Van Den~Oord, Oriol Vinyals, et~al.
\newblock Neural discrete representation learning.
\newblock \emph{Advances in neural information processing systems}, 30, 2017.

\bibitem[Chung et~al.(2021)Chung, Zhang, Han, Chiu, Qin, Pang, and Wu]{chung2021w2v}
Yu-An Chung, Yu~Zhang, Wei Han, Chung-Cheng Chiu, James Qin, Ruoming Pang, and Yonghui Wu.
\newblock W2v-bert: Combining contrastive learning and masked language modeling for self-supervised speech pre-training.
\newblock In \emph{2021 IEEE Automatic Speech Recognition and Understanding Workshop (ASRU)}, pages 244--250. IEEE, 2021.

\bibitem[Baevski et~al.(2020)Baevski, Zhou, Mohamed, and Auli]{baevski2020wav2vec}
Alexei Baevski, Yuhao Zhou, Abdelrahman Mohamed, and Michael Auli.
\newblock wav2vec 2.0: A framework for self-supervised learning of speech representations.
\newblock \emph{Advances in neural information processing systems}, 33:\penalty0 12449--12460, 2020.

\bibitem[Gulati et~al.(2020)Gulati, Qin, Chiu, Parmar, Zhang, Yu, Han, Wang, Zhang, Wu, et~al.]{gulati2020conformer}
Anmol Gulati, James Qin, Chung-Cheng Chiu, Niki Parmar, Yu~Zhang, Jiahui Yu, Wei Han, Shibo Wang, Zhengdong Zhang, Yonghui Wu, et~al.
\newblock Conformer: Convolution-augmented transformer for speech recognition.
\newblock \emph{arXiv preprint arXiv:2005.08100}, 2020.

\bibitem[Chiu et~al.(2022)Chiu, Qin, Zhang, Yu, and Wu]{chiu2022self}
Chung-Cheng Chiu, James Qin, Yu~Zhang, Jiahui Yu, and Yonghui Wu.
\newblock Self-supervised learning with random-projection quantizer for speech recognition.
\newblock In \emph{International Conference on Machine Learning}, pages 3915--3924. PMLR, 2022.

\bibitem[Ye et~al.(2024)Ye, Sun, Lei, Lin, Tan, Dai, Kong, Chen, Pan, Liu, et~al.]{ye2024codec}
Zhen Ye, Peiwen Sun, Jiahe Lei, Hongzhan Lin, Xu~Tan, Zheqi Dai, Qiuqiang Kong, Jianyi Chen, Jiahao Pan, Qifeng Liu, et~al.
\newblock Codec does matter: Exploring the semantic shortcoming of codec for audio language model.
\newblock \emph{arXiv preprint arXiv:2408.17175}, 2024.

\bibitem[Salimans and Ho(2022)]{salimans2022progressive}
Tim Salimans and Jonathan Ho.
\newblock Progressive distillation for fast sampling of diffusion models.
\newblock \emph{arXiv preprint arXiv:2202.00512}, 2022.

\bibitem[Huang et~al.(2022)Huang, Jansen, Lee, Ganti, Li, and Ellis]{huang2022mulan}
Qingqing Huang, Aren Jansen, Joonseok Lee, Ravi Ganti, Judith~Yue Li, and Daniel~PW Ellis.
\newblock Mulan: A joint embedding of music audio and natural language.
\newblock \emph{arXiv preprint arXiv:2208.12415}, 2022.

\bibitem[Lerch(2022)]{lerch2022introduction}
Alexander Lerch.
\newblock \emph{An introduction to audio content analysis: Music Information Retrieval tasks and applications}.
\newblock John Wiley \& Sons, 2022.

\bibitem[Devlin et~al.(2019)Devlin, Chang, Lee, and Toutanova]{devlin2019bert}
Jacob Devlin, Ming-Wei Chang, Kenton Lee, and Kristina Toutanova.
\newblock Bert: Pre-training of deep bidirectional transformers for language understanding.
\newblock In \emph{Proceedings of the 2019 conference of the North American chapter of the association for computational linguistics: human language technologies, volume 1 (long and short papers)}, pages 4171--4186, 2019.

\bibitem[Bartel(2006)]{bartel2006musical}
Christopher Bartel.
\newblock Musical thought and compositionality.
\newblock 2006.

\bibitem[Oord et~al.(2018)Oord, Li, and Vinyals]{oord2018representation}
Aaron van~den Oord, Yazhe Li, and Oriol Vinyals.
\newblock Representation learning with contrastive predictive coding.
\newblock \emph{arXiv preprint arXiv:1807.03748}, 2018.

\bibitem[Du et~al.(2025)Du, Yang, and Welleck]{du2025optimizing}
Weihua Du, Yiming Yang, and Sean Welleck.
\newblock Optimizing temperature for language models with multi-sample inference.
\newblock \emph{arXiv preprint arXiv:2502.05234}, 2025.

\bibitem[Ho and Salimans(2022)]{ho2022classifier}
Jonathan Ho and Tim Salimans.
\newblock Classifier-free diffusion guidance.
\newblock \emph{arXiv preprint arXiv:2207.12598}, 2022.

\bibitem[Kreuk et~al.(2022)Kreuk, Synnaeve, Polyak, Singer, D{\'e}fossez, Copet, Parikh, Taigman, and Adi]{kreuk2022audiogen}
Felix Kreuk, Gabriel Synnaeve, Adam Polyak, Uriel Singer, Alexandre D{\'e}fossez, Jade Copet, Devi Parikh, Yaniv Taigman, and Yossi Adi.
\newblock Audiogen: Textually guided audio generation.
\newblock \emph{arXiv preprint arXiv:2209.15352}, 2022.

\bibitem[Touvron et~al.(2023)Touvron, Lavril, Izacard, Martinet, Lachaux, Lacroix, Rozi{\`e}re, Goyal, Hambro, Azhar, Rodriguez, Joulin, Grave, and Lample]{touvron2023llama}
Hugo Touvron, Thibaut Lavril, Gautier Izacard, Xavier Martinet, Marie-Anne Lachaux, Timoth{\'e}e Lacroix, Baptiste Rozi{\`e}re, Naman Goyal, Eric Hambro, Faisal Azhar, Aurelien Rodriguez, Armand Joulin, Edouard Grave, and Guillaume Lample.
\newblock Llama: Open and efficient foundation language models.
\newblock \emph{arXiv preprint arXiv:2302.13971}, 2023.

\bibitem[Graves et~al.(2006)Graves, Fern{\'a}ndez, Gomez, and Schmidhuber]{graves2006connectionist}
Alex Graves, Santiago Fern{\'a}ndez, Faustino Gomez, and J{\"u}rgen Schmidhuber.
\newblock Connectionist temporal classification: labelling unsegmented sequence data with recurrent neural networks.
\newblock In \emph{Proceedings of the 23rd international conference on Machine learning}, pages 369--376, 2006.

\bibitem[Zhang et~al.(2023{\natexlab{b}})Zhang, Han, Qin, Wang, Bapna, Chen, Chen, Li, Axelrod, Wang, et~al.]{zhang2023google}
Yu~Zhang, Wei Han, James Qin, Yongqiang Wang, Ankur Bapna, Zhehuai Chen, Nanxin Chen, Bo~Li, Vera Axelrod, Gary Wang, et~al.
\newblock Google usm: Scaling automatic speech recognition beyond 100 languages.
\newblock \emph{arXiv preprint arXiv:2303.01037}, 2023{\natexlab{b}}.

\bibitem[Kong and Ping(2021)]{kong2021fast}
Zhifeng Kong and Wei Ping.
\newblock On fast sampling of diffusion probabilistic models.
\newblock In \emph{ICML Workshop on Invertible Neural Networks, Normalizing Flows, and Explicit Likelihood Models}, 2021.

\bibitem[Lanzend{\"o}rfer et~al.(2023)Lanzend{\"o}rfer, Gr{\"o}tschla, Funke, and Wattenhofer]{lanzendorfer2023disco}
Luca Lanzend{\"o}rfer, Florian Gr{\"o}tschla, Emil Funke, and Roger Wattenhofer.
\newblock Disco-10m: A large-scale music dataset.
\newblock \emph{Advances in Neural Information Processing Systems}, 36:\penalty0 54451--54471, 2023.

\bibitem[D{\'e}fossez et~al.(2019)D{\'e}fossez, Usunier, Bottou, and Bach]{defossez2019demucs}
Alexandre D{\'e}fossez, Nicolas Usunier, L{\'e}on Bottou, and Francis Bach.
\newblock Demucs: Deep extractor for music sources with extra unlabeled data remixed.
\newblock \emph{arXiv preprint arXiv:1909.01174}, 2019.

\bibitem[D{\'e}fossez(2021)]{defossez2021hybrid}
Alexandre D{\'e}fossez.
\newblock Hybrid spectrogram and waveform source separation.
\newblock In \emph{Proceedings of the ISMIR 2021 Workshop on Music Source Separation}, 2021.

\bibitem[Rouard et~al.(2023)Rouard, Massa, and D{\'e}fossez]{rouard2022hybrid}
Simon Rouard, Francisco Massa, and Alexandre D{\'e}fossez.
\newblock Hybrid transformers for music source separation.
\newblock In \emph{ICASSP 23}, 2023.

\bibitem[OpenAI(2023)]{chatgpt}
OpenAI.
\newblock Chatgpt.
\newblock \emph{URL https://chat.openai.com/}, 2023.

\bibitem[team(2024{\natexlab{a}})]{sunov4}
Suno team.
\newblock Introducing v4.
\newblock \emph{URL https://suno.com/blog/v4}, 2024{\natexlab{a}}.

\bibitem[team(2024{\natexlab{b}})]{udio1.5}
Udio team.
\newblock Introducing v1.5.
\newblock \emph{URL https://www.udio.com/blog/introducing-v1-5}, 2024{\natexlab{b}}.

\bibitem[team(2024{\natexlab{c}})]{mureka5.5}
Mureka team.
\newblock Mureka ai.
\newblock \emph{URL https://www.mureka.ai}, 2024{\natexlab{c}}.

\bibitem[Kilgour et~al.(2019)Kilgour, Zuluaga, Roblek, and Sharifi]{kilgour2019frechet}
Kevin Kilgour, Mauricio Zuluaga, Dominik Roblek, and Matthew Sharifi.
\newblock Fr{\'e}chet audio distance: A reference-free metric for evaluating music enhancement algorithms.
\newblock In \emph{INTERSPEECH}, pages 2350--2354, 2019.

\bibitem[Rafii et~al.(2019)Rafii, Liutkus, St{\"o}ter, Mimilakis, and Bittner]{rafii2019musdb18}
Zafar Rafii, Antoine Liutkus, Fabian-Robert St{\"o}ter, Stylianos~Ioannis Mimilakis, and Rachel Bittner.
\newblock Musdb18-hq-an uncompressed version of musdb18.
\newblock \emph{(No Title)}, 2019.

\bibitem[Gemmeke et~al.(2017)Gemmeke, Ellis, Freedman, Jansen, Lawrence, Moore, Plakal, and Ritter]{gemmeke2017audio}
Jort~F Gemmeke, Daniel~PW Ellis, Dylan Freedman, Aren Jansen, Wade Lawrence, R~Channing Moore, Manoj Plakal, and Marvin Ritter.
\newblock Audio set: An ontology and human-labeled dataset for audio events.
\newblock In \emph{2017 IEEE international conference on acoustics, speech and signal processing (ICASSP)}, pages 776--780. IEEE, 2017.

\bibitem[Tjandra et~al.(2025)Tjandra, Wu, Guo, Hoffman, Ellis, Vyas, Shi, Chen, Le, Zacharov, et~al.]{tjandra2025meta}
Andros Tjandra, Yi-Chiao Wu, Baishan Guo, John Hoffman, Brian Ellis, Apoorv Vyas, Bowen Shi, Sanyuan Chen, Matt Le, Nick Zacharov, et~al.
\newblock Meta audiobox aesthetics: Unified automatic quality assessment for speech, music, and sound.
\newblock \emph{arXiv preprint arXiv:2502.05139}, 2025.

\end{thebibliography}


\end{document}